\documentclass[a4paper]{jpconf}
\usepackage{graphicx}
\usepackage{epsfig}
\usepackage{latexsym}
\usepackage{amssymb}
\usepackage{amsbsy}
\usepackage{amsmath}
\usepackage{cite}
\usepackage{epsfig}
\usepackage{graphics,color}
\usepackage{a4}

\newcommand{\be}{\begin{equation}}
\newcommand{\ee}{\end{equation}}

\newcommand{\bea}{\begin{eqnarray}}
\newcommand{\eea}{\end{eqnarray}}
\newcommand{\bean}{\begin{eqnarray*}}
\newcommand{\eean}{\end{eqnarray*}}

\def\Etmisscal{\hbox{$\rlap{\kern0.25em/}E_T^{cal}$}}
\def\Etmiss{\hbox{$\rlap{\kern0.25em/}E_T$}}

\def \Bul {\ifmmode {\bullet }\else {$\bullet $\enskip}\fi}
\def \Ebul {\ifmmode {\circ }\else {$\circ $\enskip}\fi}
\def \imply {\ifmmode {\Rightarrow }\else {$\Rightarrow $}\fi}

\def \Gsim {$ \sim$\llap {$^>$}}

\def\Ang{
\setbox0=\hbox{h}\dimen1=\ht0 \advance\dimen1 by-1ex
\rlap{\raise.67\dimen1\hbox{{}\hskip2pt\char'27}}A}
\def \Bul {~~$\bullet $~}

\def\tt{\mbox{$t\bar{t}$}}

\def\JNX#1#2{JET\_#1\_#2}

\def\bsds3pi{\mbox{$B_s \rightarrow D_s + 3\pi$}}

\def\ds3pi{\mbox{$D_s + 3\pi$}}

\def \imply {\ifmmode {\Rightarrow }\else {$\Rightarrow $}\fi}

\def\bsds3pi{\mbox{$B_s^{\circ} \rightarrow D_s^- + 3\pi^{\pm}$}}

\def\ds3pi{\mbox{$D_s^- + 3\pi^{\pm}$}}

\def\JNX#1#2{JET\_#1\_#2}
\def\J3M{\JNX{3}{MED}}

\def\JNX#1#2{JET\_#1\_#2}
\def\J3M{\JNX{3}{MED}}

\begin{document}
\title{Making the most of \\ Taylor expansion and imaginary $\mu$ }

\author{E.~Laermann, F.~Meyer} 

\address{Fakult\"at f\"ur Physik, Universit\"at Bielefeld,D--33615 Bielefeld, Germany}

\author{M.~P.~Lombardo}
\address{INFN-Laboratori Nazionali di Frascati, I--00044, Frascati (RM) Italy}

\begin{abstract}
We present preliminary results for the curvature of the
pseudocritical line and susceptibilities in $N_f = 2 + 1$ 
flavor QCD. 
The computations are carried out on lattice sizes of $16^3 \times 4$, at 
matching parameters of early  work of the Bielefeld group. 
Emphasis is placed on the control of systematic errors, by 
cross-validating results obtained by use of the Taylor expansion and
measurements at imaginary chemical potential. To this end,
we generalize the magnetic equation of state to the analysis
of the number density, and we extend it to 
imaginary values of the chemical potential. 
\end{abstract}

\section{Introduction}
Most likely, a large fraction of the  participants in this meeting are
familiar with the challenges faced by lattice QCD  at non-zero
baryon density\cite{Aarts}: the region of quark chemical potential $\mu_q$
exceeding $m_\pi/2 $ remains prohibitively difficult, due to a  
severe sign problem\cite{Splittorff}.  

For smaller chemical potential a variety of methods have been proposed 
to circumvent these difficulties, which have been quite successful 
in many respects. However, even in the easy region $\mu_q < m_\pi/2$ 
there are challenges to face: for instance, it has been noted that most current 
estimates of the QCD endpoint lie dangerously close
to the 'forbidden region' $\mu_q > m_\pi/2$; the 
qualitative agreement among different results on the slope of the
critical line still needs to be corroborated by some convincing continuum
estimates obtained with different methods. And, there is a growing interest
in the values of higher order susceptibilities, particularly
at larger values of the chemical potential, while the current
results are limited to the lowest order ones \cite{Lombardo}. So there is
room for improvement, which can perhaps be achieved 
by  mixing and matching different approaches. 

\section{Two complemetary approaches: Taylor expansion and imaginary chemical
potential}
One  way to handle the physics of small chemical potential
relies on Taylor expansion\cite{Cheng}. 
This approach is well understood mathematically,
however it only works inside its convergency
circle, whose radius is unknown. Moreover even inside the circle
it is hard to judge  when
convergence has been achieved. Alternatively,
analytic continuation from imaginary chemical potential -- where
conventional simulations are possible --  to real 
chemical potential  can in principle be extended to  the entire 
analytic domain, but the danger  here is a possible lack of control over
the continuation itself: for instance, terms which are subleading for
imaginary chemical potential might well become leading 
for real chemical potential. The strengths of the two methods are somewhat
complementary, and their weaknesses different: this suggests to develop 
a hybrid method  which combines the two, and  this note reports on our
ongoing effort in this direction\cite{Meyer}.
\begin{figure}
\vskip -12 truecm
\begin{minipage}{21truecm}
\hskip -2.0 truecm \includegraphics[width=9truecm]{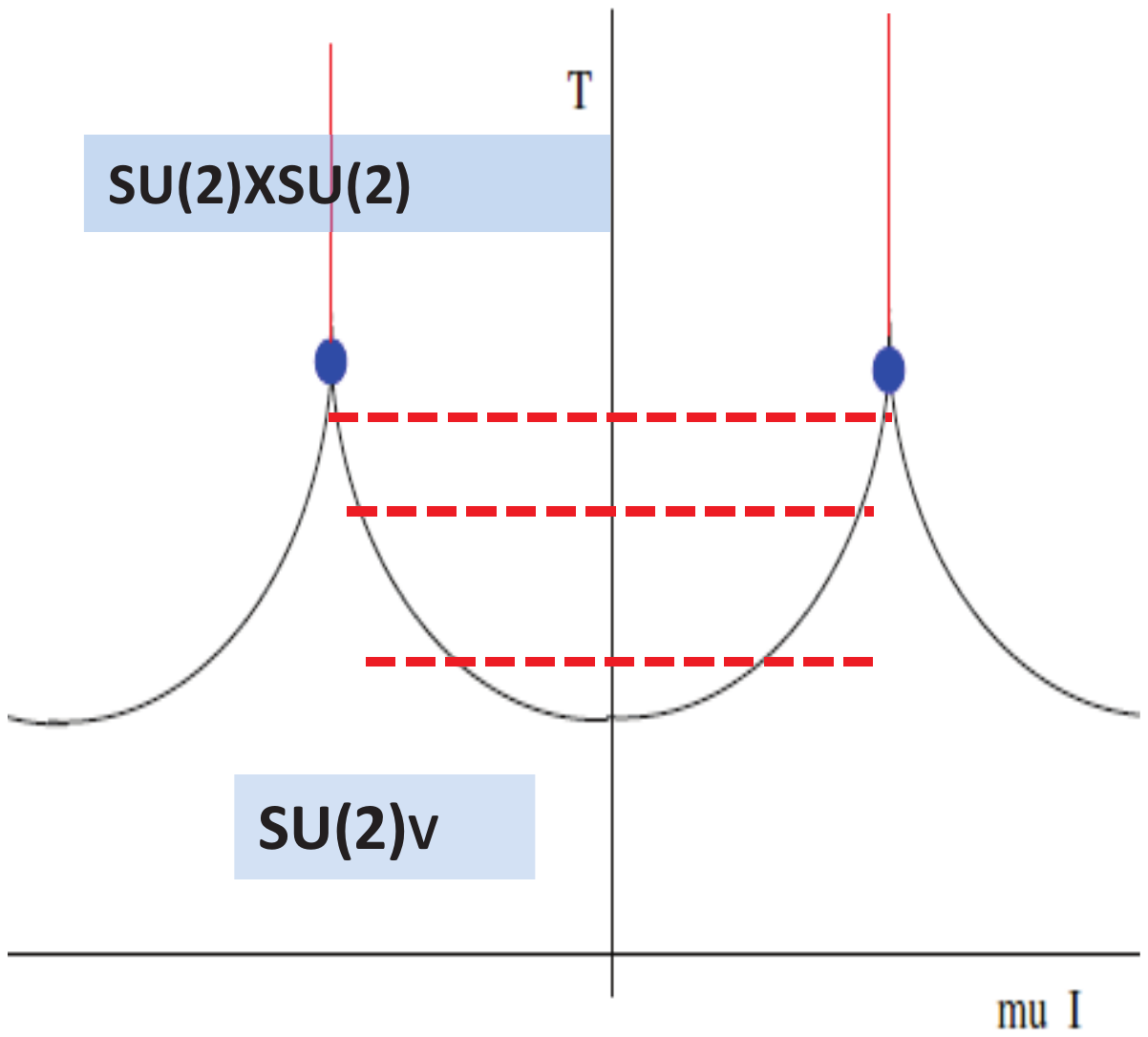}\hskip-2.0truecm\includegraphics[width=12truecm]{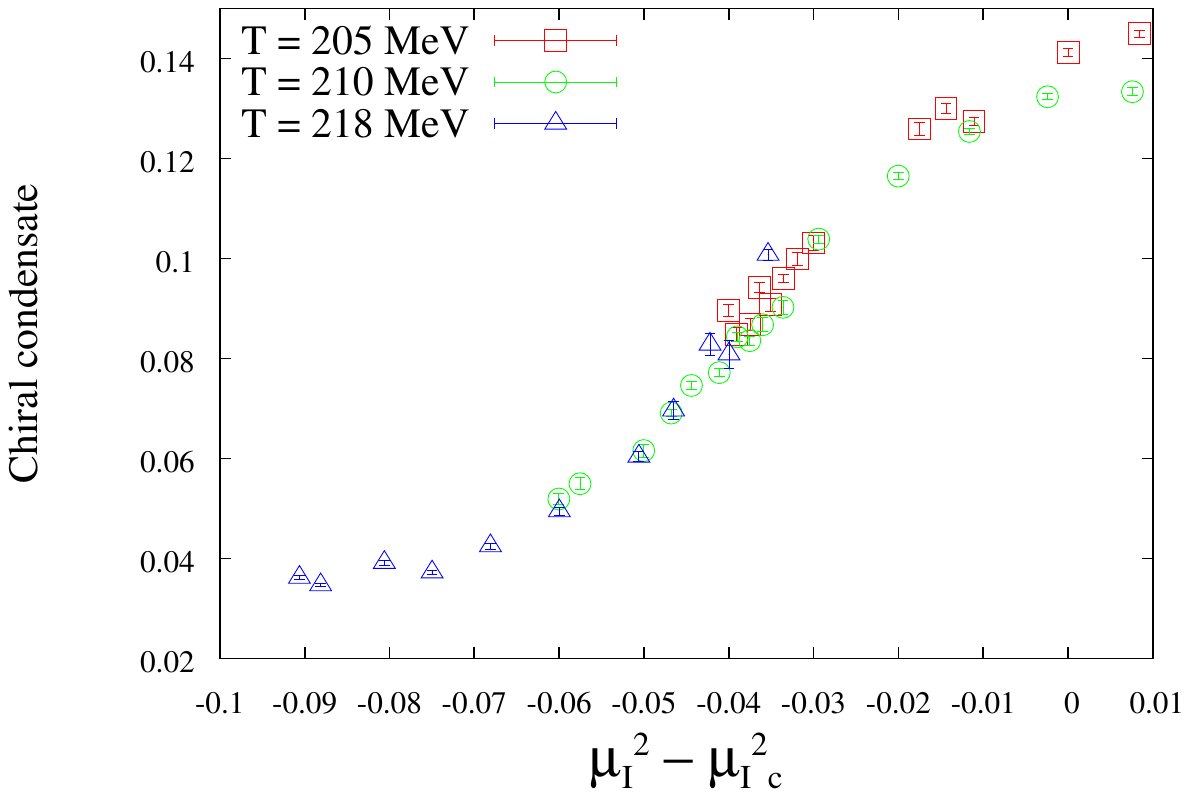}
\end{minipage}
\vskip -1.0 truecm
\caption{A sketchy view of the phase diagram of QCD 
for two massless flavors in the temperature, imaginary 
chemical potential plane, with the superimposed  trajectories
of our simulations. The blue points indicate 
the Roberge-Weiss endpoint (left). The chiral condensate measured along
such trajectories,  plotted against the rescaled chemical 
potential $\mu^2 - \mu^2_c(T)$ (right). Note
the (approximate) universal behaviour of the 
condensate. } 
\end{figure}

\section{Results}    
 Our study for staggered 2+1 flavors on lattices of size $16^3 \times 4$ clearly is exploratory. However, the quark mass values are close to the ones realized in nature. The Goldstone pion mass is tuned to about $220$ MeV and the kaon acquires its physical mass. This corresponds to a (degenerate) light to strange quark mass ratio of $m_l /m_s = 1/10$. The action utilized is the p4fat3 action for which Taylor coefficients computed at $\mu = 0$ are available in our  range 
of temperatures and at our quark masses\cite{Cheng}.
The computations were carried out at three temperatures above 
the critical temperature for two massless flavors
for the considered lattice spacings: $T = 205, 210, 218$ MeV. 
In Fig.1, left we sketch
the simulations' trajectories, alongside the critical line in the chiral limit,
in the $\mu_I, T$ plane. The
sketch evidences  the $\mu_I$ periodicity as well as the symmetry under
reflection $\mu_I \to - \mu_I$; the actual simulations were performed only at positive three-flavor degenerate 
$\mu_I$, and within the first half--period.
The blue points mark the endpoint of the Roberge Weiss phase transition. 
The light quark chiral condensate, plotted in Fig.1, right,
 is showing a fairly rapid rise when 
$\mu_I$ is increased,  signaling
the pseudocritical point for the transition 
to the broken phase (marked $SU(2)_V$ in the diagram).
  We plot the results against a rescaled chemical
potential  to confirm the feasibility of a scaling study
based on the magnetic Equation of
State. 

\subsection{The magnetic Equation of State}

Consider now the magnetic equation of state
\begin{equation}
M =  h^{1/\delta} f_G(z) \hskip 1 truecm  z = {t}{h^{\frac{-1}{\beta \delta}}} 
\end{equation}
where $M$ and $h$ are identified with the light chiral condensate and the
light bare mass: $M = m_s \langle \bar \psi \psi \rangle_l$,
$h = \frac{1}{h_0} \frac{m_l}{m_s}$ respectively, 
$f_G$ is a scaling function which only depends
on the universality class of the transition, and  
$t=0$ defines the critical line. 
Our light mass does not change:
if scaling (i.e. Eq. 1) holds true the chiral condensate results plotted 
versus $\mu^2 - \mu^2_c(T)$, where $\mu^2_c(T)$
is the critical chemical potential for each temperature $T$, should fall
on a single curve. In Fig.1 we see that this is indeed happening,
 demonstrating the universal behaviour of the chiral
condensate at least in the range of parameters we are exploring.
$\mu^2_c(T)$ is found by searching for the optimal
match between results for different temperatures, and compares
favorably with alternative estimates. 

\subsection{Curvature of the critical line from the magnetic EoS}
For small chemical potential   $t$ can be explicitely written as 
\begin{equation}
t = \frac {1}{t_0} \left (\frac{T - T_c}{T_c} +  k ( \mu / T) ^2 \right)   
\end{equation}
where $T_c$ is the critical temperature in the chiral limit.
The standard condition $f_G(0) = f'_G(0) = 1$ fixes the normalization $t_0$ and 
$h_0$ which have been determined in Ref. \cite{Ejiri}. As a working hypothesis
we assume that  $t_0$ and $h_0$ do not depend on the temperature, and
we  check a posteriori that this indeed is the case.
$k$ -- the curvature of the critical line -- can then be computed from the
Equation of State, using the explicit forms  available for the 
two plausible universality classes $O(2)$ and $O(4)$, and the results
contrasted with those obtained in Ref.\cite{Kaczmarek}.

\begin{figure}[t]
\vskip-10 truecm
\begin{minipage}{21truecm}
\hskip -0.8truecm
\includegraphics[width=7.5truecm,height=13truecm]{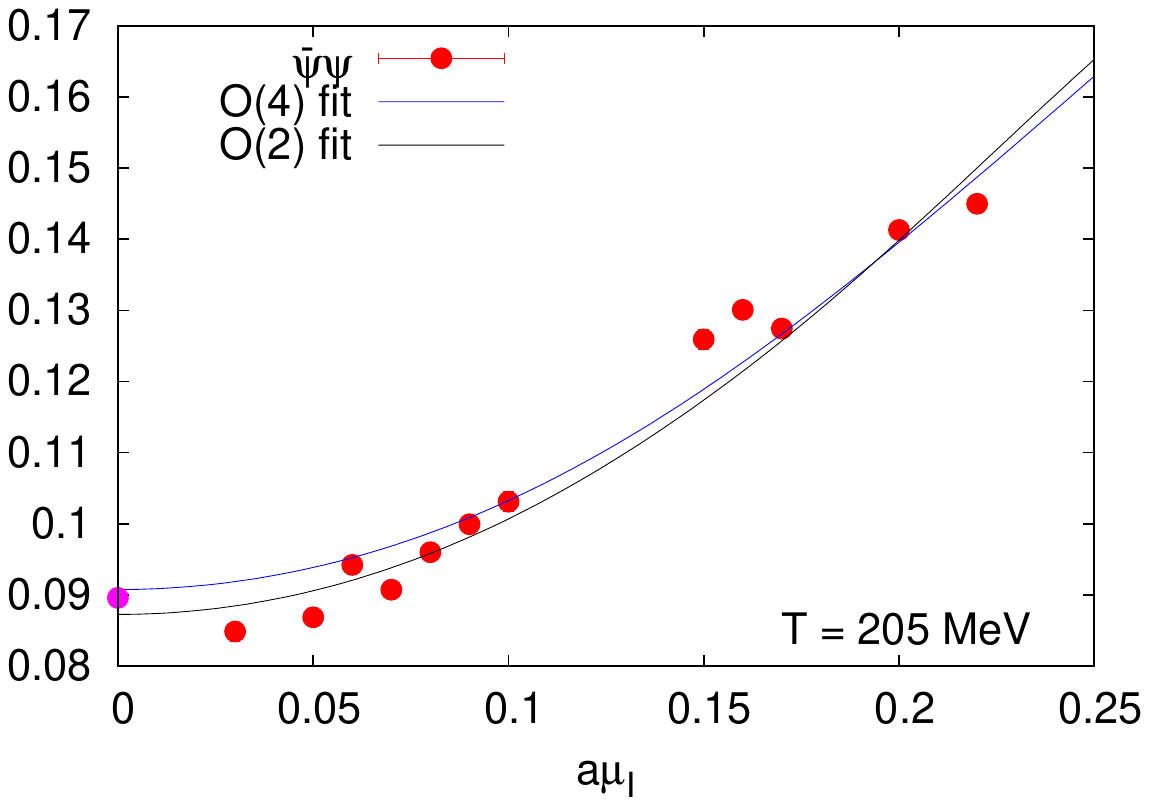}
\hskip -3.1 truecm
\includegraphics[width=7.5truecm,height=13truecm]{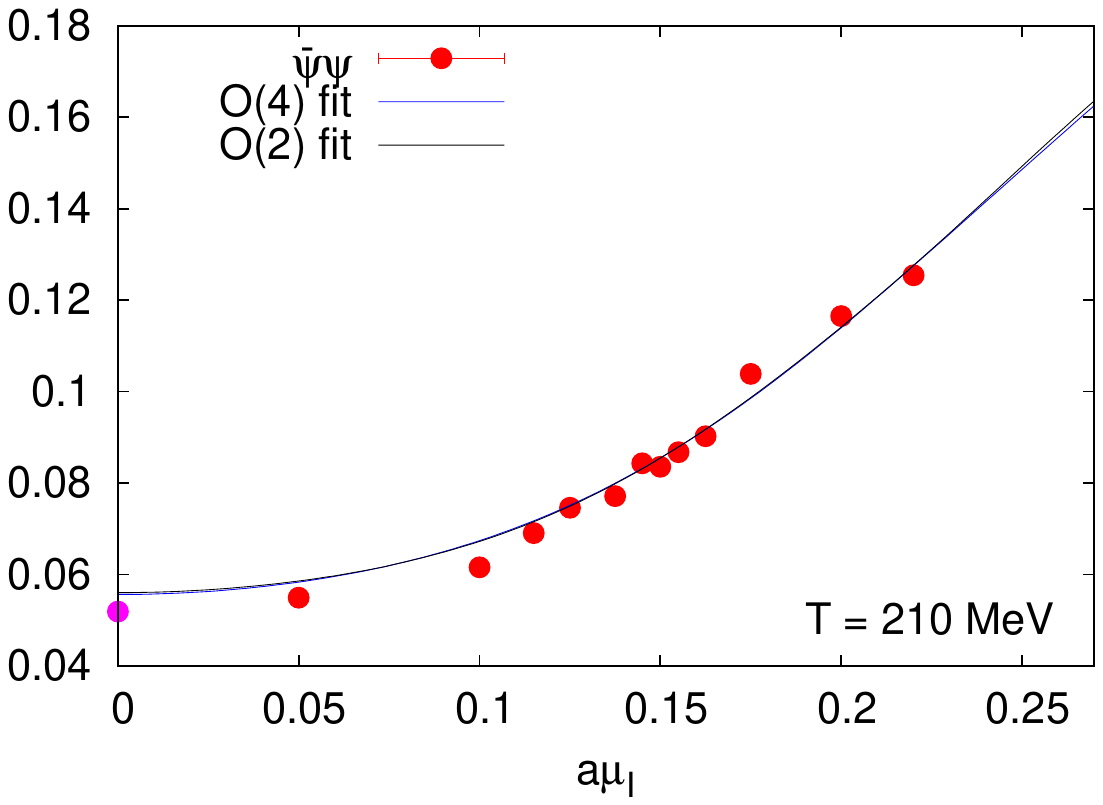}
\hskip -3.1 truecm
\includegraphics[width=7.5truecm,height=13truecm]{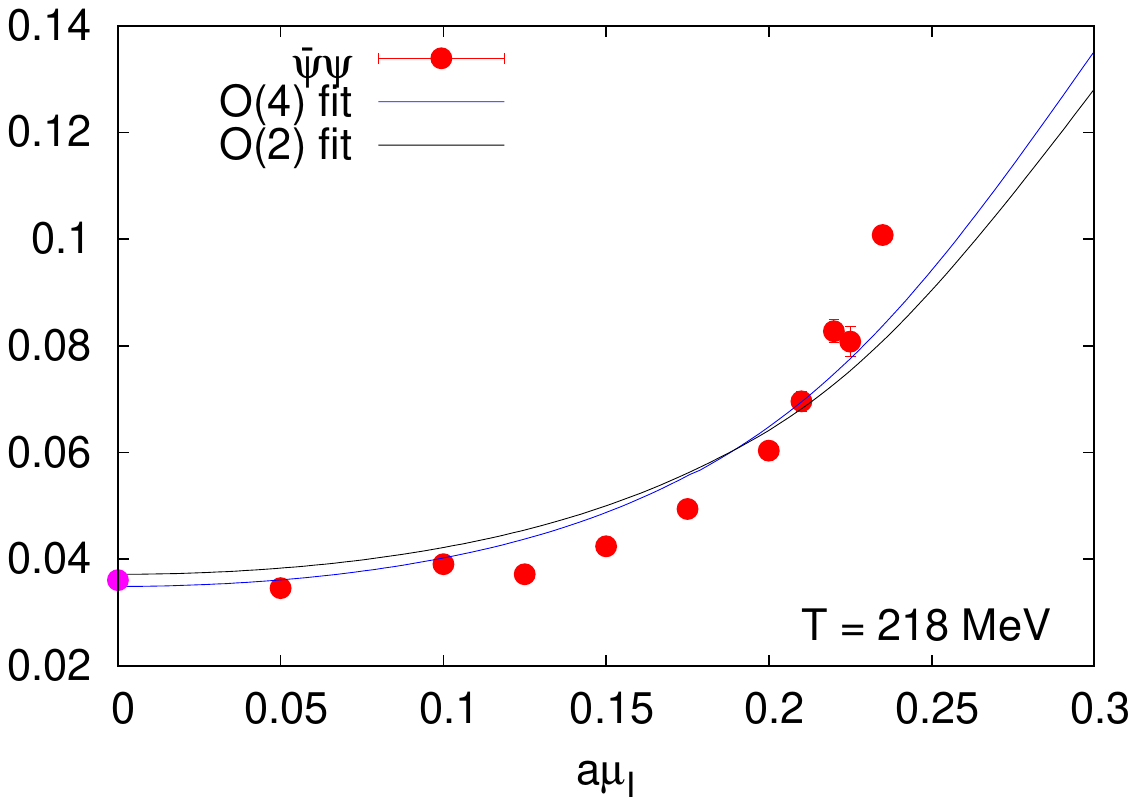}
\end{minipage}
\vskip -1truecm
\caption{The fits motivated by the magnetic Equation of State for
the chiral condensate,superimposed to the data point.}
\end{figure}

\begin{table}
\caption{The results for the curvature of the critical line 
$k$ (Eq.(2)) from the fits to the magnetic EoS for the
chiral condensate ($k_{\bar \psi \psi}$), and from the fits to the  number density ($k_n$).}
\begin{tabular}{|c||c c c || c c c | }
\hline
& & $k_{\bar \psi \psi}$  & &  & $k_n$  &  \\
\hline
T(MeV) & 205  & 210  & 218 &   205  & 210  & 218  \\
\hline
O(2)  &  0.026(1) & 0.031(1) & 0.025(3) & 0.021(1) & 0.021(1)& 0.016(1) \\                             
O(4)  & 0.022(1)  & 0.028(1) & 0.025(3) & 0.039(1) & 0.036(1) & 0.023(2)  \\
\hline
\end{tabular}
\end{table}
From the EoS for the order parameter 
one can derive analogous relations for the magnetic susceptibility,
as well as for the number density $n$ and its susceptibility\cite{Engels},  
leading to alternative  estimates of $k$. To fix the notation,
let us   write  the Taylor 
expansion of the pressure in terms of the quark chemical potentials
around a generic $\mu_0 = (\mu_{u,0}, \mu_{d,0}, \mu_{s,0})$
for our $N_f = 2+1$ system:
\begin{equation}
  {p\over T^4}(\hat{\mu})=\sum_{k,l,n}c_{kln}(\hat{\mu}_u - \hat{\mu}_{u,0})^k
(\hat{\mu}_d - \hat{\mu}_{d,0})^l(\hat{\mu}_s - \hat{\mu}_{s,0})^n\nonumber
\end{equation}
with the abbreviation $\hat{\mu}_q = \mu_q/T$ and
 the coefficients
\begin{equation}
  c_{kln} = \frac{1}{k!l!n!}\frac{\partial^k}{\partial\hat{\mu_u}^k}
\frac{\partial^l}{\partial\hat{\mu_d}^l}
\frac{\partial^n}{\partial\hat{\mu_s}^n}\left(p \over T^4\right)
\end{equation}
evaluated at $\mu_0$.
We will in particular discuss 
$c_1 = c_{100}$ and $c_2 = c_{200}$, which are related
to the particle number and its susceptibility. 
In Fig.2 and Fig.3 we show
the results of the EoS fits  and in  Table 1
we record the $k$ values returned by the fits for the
chiral condensate and the quark number (the coefficient $c_1$).
While the fit results, at least for O(2), do not seem to be inconsistent, it 
is clear that at values for $\mu_I$ close to the Roberge-Weiss limit 
other scaling studies need to be performed, work for the future.
\begin{figure}
\vskip-7.5 truecm
\begin{minipage}{21truecm}
\hskip -0.8truecm
\includegraphics[width=7.5truecm,height=13truecm]{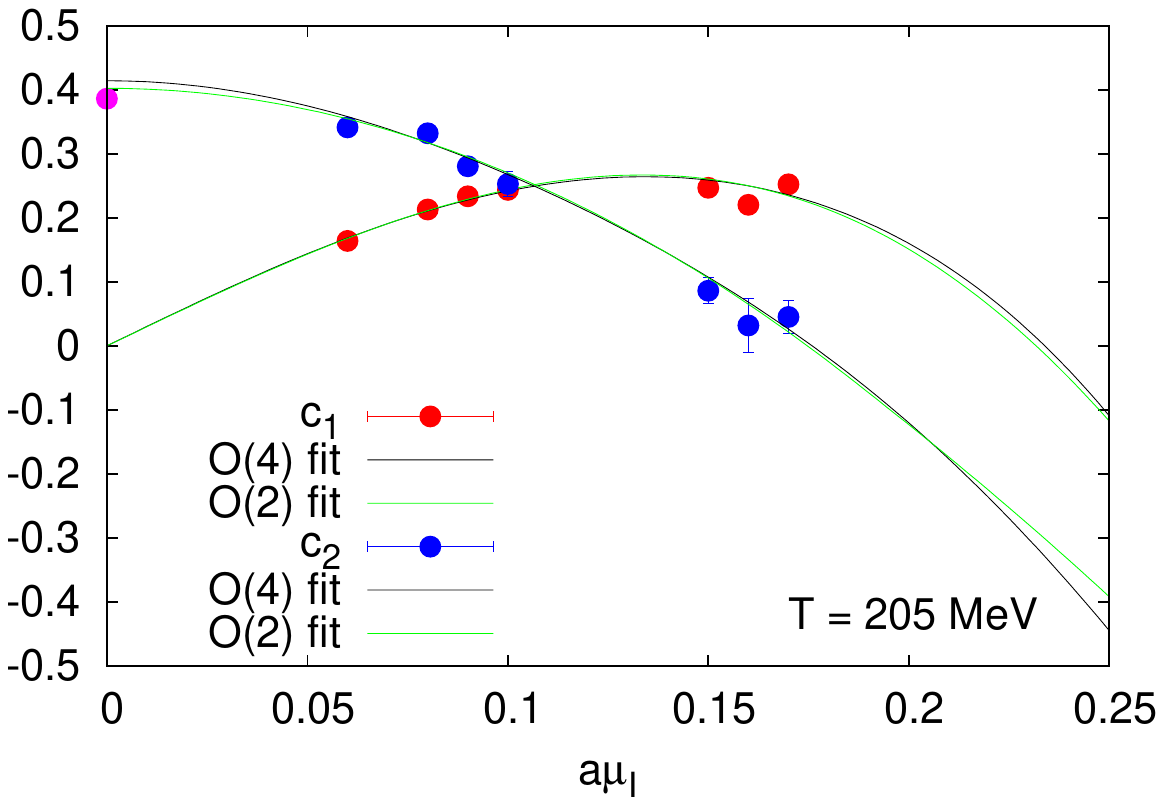}
\hskip -3.1 truecm
\includegraphics[width=7.5truecm,height=13truecm]{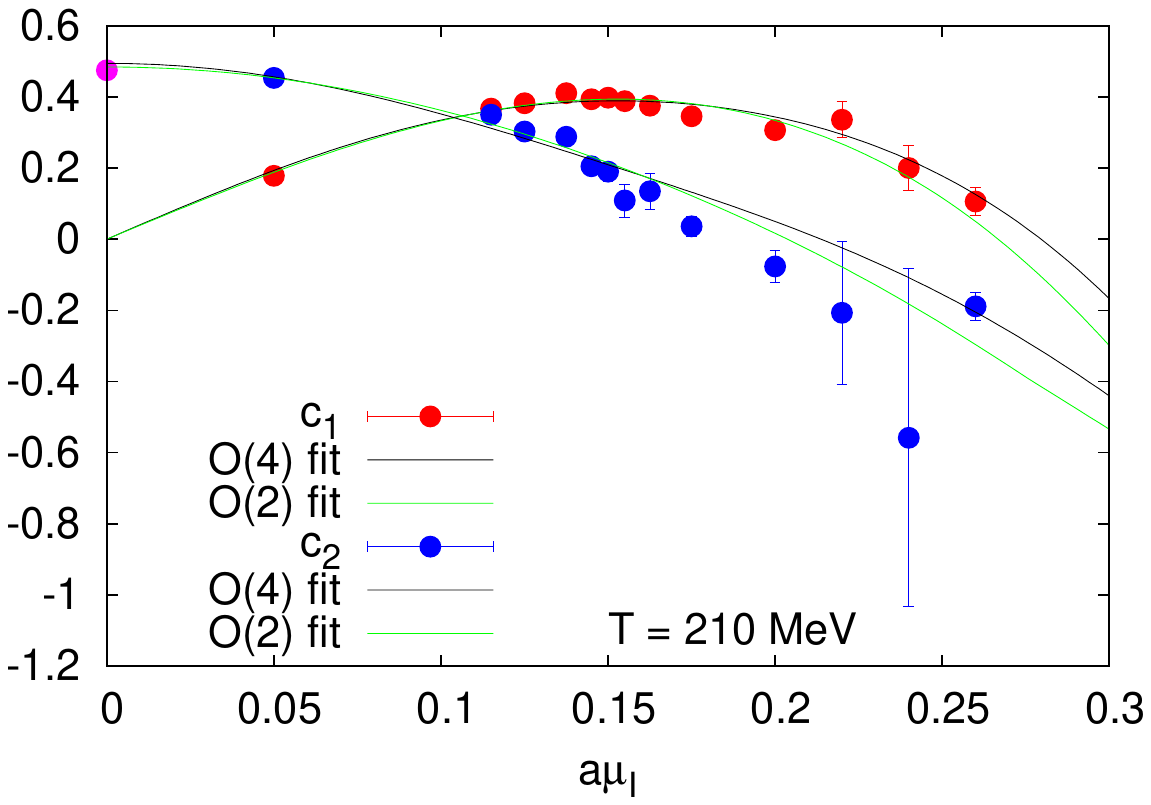}
\hskip -3.1 truecm
\includegraphics[width=7.5truecm,height=13truecm]{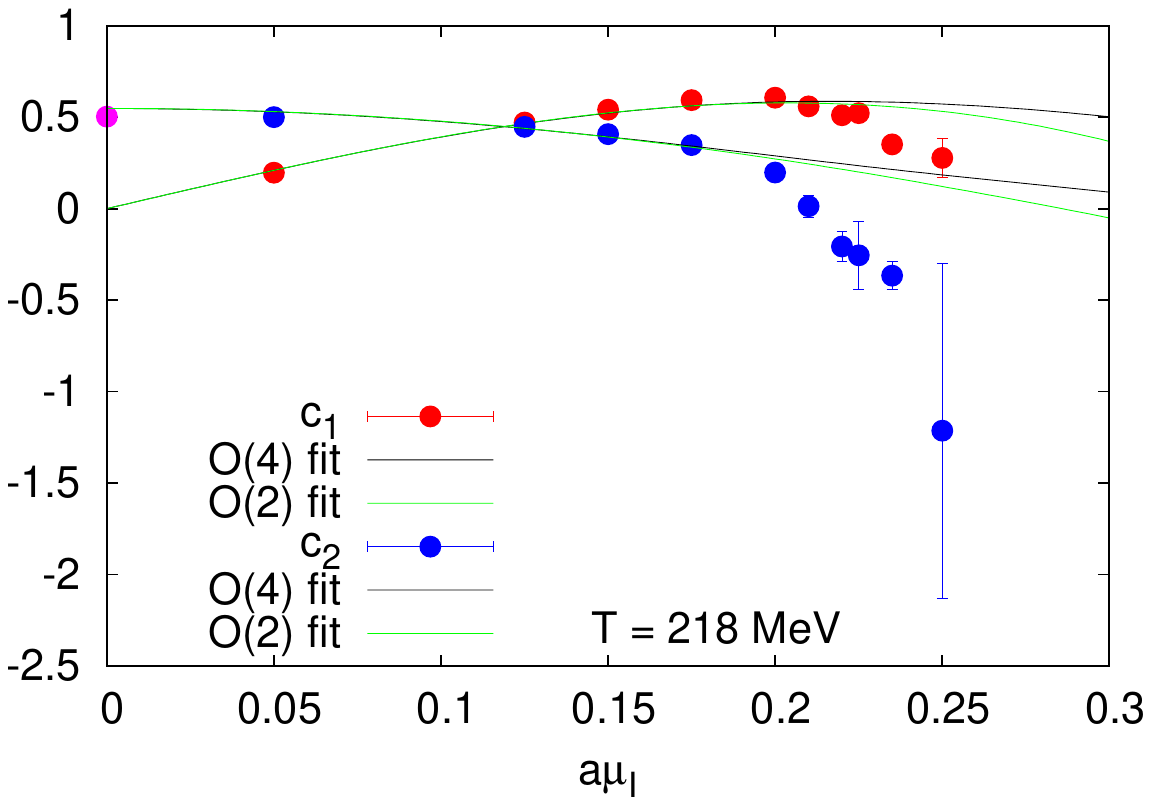}
\end{minipage}
\vskip -1truecm
\caption{The fits motivated by the magnetic Equation of State for
 $c_1$  and $c_2$ (related to quark number and 
its susceptibility respectively, see text) superimposed to the data points.}
\end{figure}

\subsection{The Taylor expansion}
How far in $\mu$ we can trust the Taylor expansion? A direct comparison
of  the number density of e.g. the u quark  given by the partial
sums of its Taylor series:
\begin{equation}
  {n_u\over T^3}(\hat{\mu})=\sum_{k,l,n}kc_{kln}
(\hat{\mu}_u - \hat{\mu}_{u,0})^{k-1}
(\hat{\mu}_d - \hat{\mu}_{d,0})^l(\hat{\mu}_s - \hat{\mu}_{s,0})^n.
\end{equation}
with results obtained directly at imaginary $\mu$ might give the answer. We show
the results for increasing temperatures in Fig.4:
for $T = 205$ MeV   a good convergence is achieved indicating a small
contribution from the $8^{th}$ order terms and superior ones. At the highest
temperature instead the results suggest that a  $6^{th}$ order expansion
is not capable to describe the data at $\mu_I$ \Gsim $0.2$. 
The simultaneous consideration of Taylor expansion and imaginary
$\mu$ results naturally suggests a strategy to extract higher order
terms of the expansions -- or, equivalently, higher order susceptibilities --
 withouth having to implement the cumbersome higher order derivatives.
At the same time one can place lower bounds on the radius of convergence
simply by assessing by eye the extent of the region where actual
convergence is achieved. 

\begin{figure}
\vskip-7.5 truecm
\begin{minipage}{21truecm}
\hskip -0.8truecm
\includegraphics[width=7.5truecm,height=13truecm]{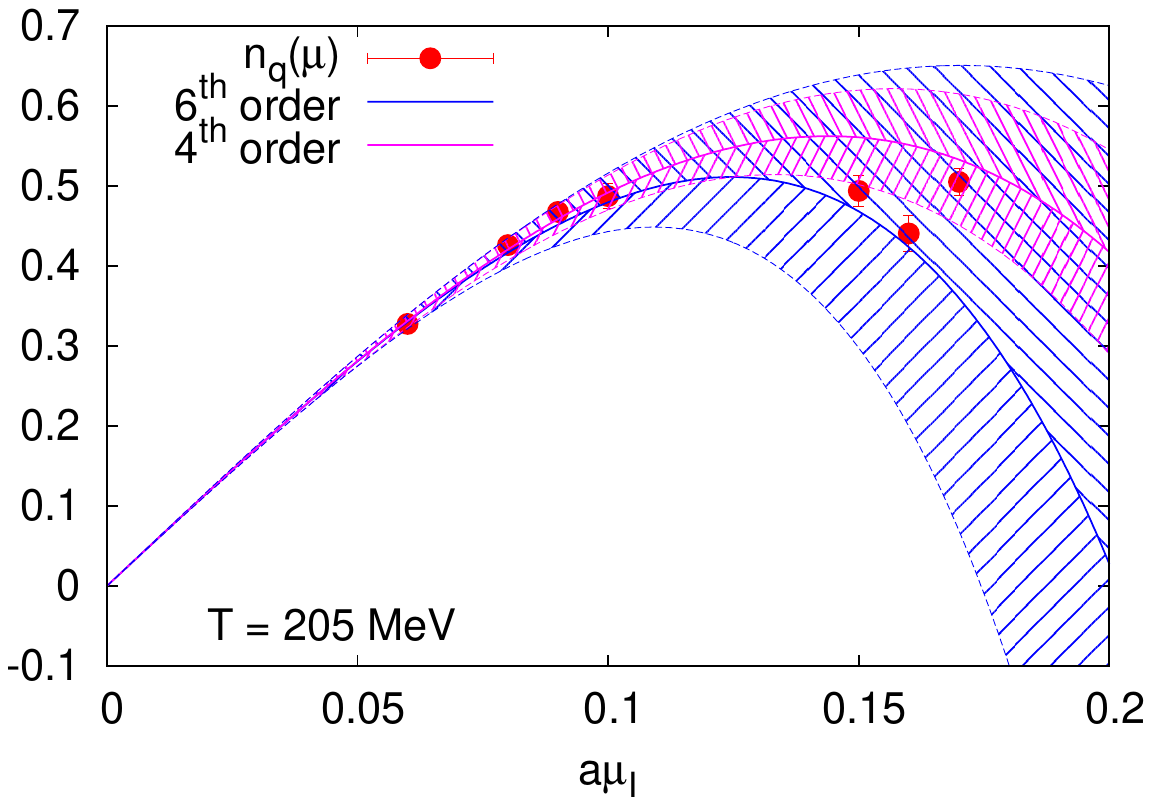}
\hskip -3.1 truecm
\includegraphics[width=7.5truecm,height=13truecm]{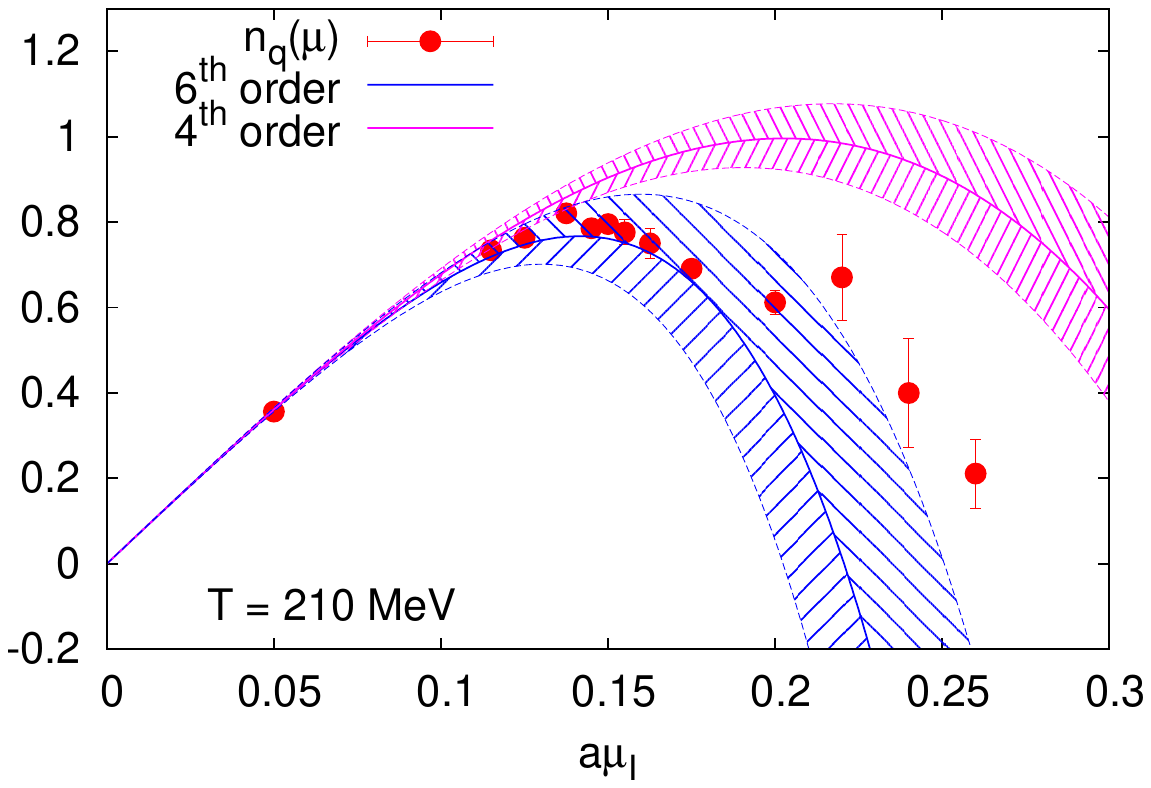}
\hskip -3.1 truecm
\includegraphics[width=7.5truecm,height=13truecm]{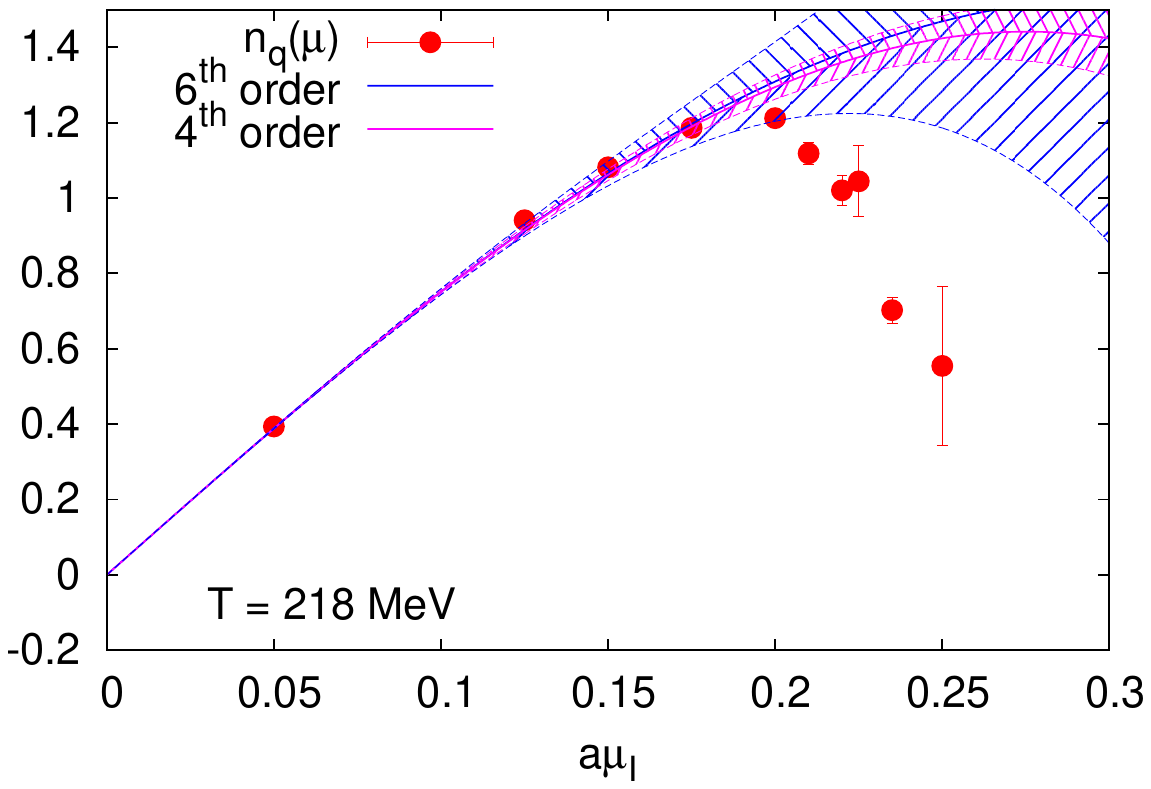}
\end{minipage}
\vskip -1truecm
\caption{The results for the number density at imaginary
chemical potential, contrasted with the partial sums
of the Taylor series of the indicated order.} 
\end{figure}

\section{Parting comments}

We have presented an overview of results obtained by
performing simulations at zero and non--zero imaginary chemical potential.
We have noted that  an analysis based on the magnetic equation of state 
for the chiral condensate, the number density
and their susceptibilities 
at imaginary chemical potential gives information on the universality 
class of the transition for two massless flavors,
and on the slope of the chiral line.  We have argued that  
the simultaneous consideration of results 
from the Taylor expansion at $\mu=0$ and at 
imaginary $\mu$ can place a lower bound on the radius of convergence 
at $\mu=0$, and facilitate the computation of higher order coefficients.

\end{document}